# Hybrid Classical-Quantum Neural Networks for Multi-Characteristic Co-Optimization of Recessed-Gate AlGaN/GaN MIS-HEMTs

Rushat Rai[1], Pei-Jie Chang[2], Doan Viet Nguyen[3], Yuan-Chieh Chiu[2], Niall Tumilty[1], Yun-Yuan Wang[4], Simon See[5], Wen-Jay Lee[6,*], Tai-Yue Li[6,*], Nan-Yow Chen[6,*], and Tian-Li Wu[1,2,3,*]

[1] International College of Semiconductor Technology, National Yang Ming Chiao Tung University, Hsinchu 300093, Taiwan

[2] Institute of Electronics, National Yang Ming Chiao Tung University, Hsinchu 300093, Taiwan

[3] EECS International Graduate Program, National Yang Ming Chiao Tung University, Hsinchu 300093, Taiwan

[4] NVIDIA AI Technology Center, NVIDIA Corp., Taipei 115004, Taiwan

[5] NVIDIA AI Technology Center, NVIDIA Corp., Singapore 038988, Singapore

[6] National Center for High-performance Computing, National Applied Research Laboratories, Hsinchu 300092, Taiwan

Corresponding author: wjlee@niar.org.tw, 2503001@niar.org.tw, nanyow@niar.org.tw, tlwu@nycu.edu.tw

## Abstract

Optimizing recessed-gate AlGaN/GaN MIS-HEMTs requires accurate multi-characteristic models, but experimental semiconductor datasets remain costly and encode process-induced variability that simulations cannot faithfully reproduce. This work proposes a hybrid classical-quantum neural network (HQNN) for joint optimization of six electrical targets from a 24-dimensional fabrication/process vector. We systematically screen quantum-circuit templates to extract circuit-design guidance, then select a final HQNN and compare it directly with classical baselines. On 468 experimentally fabricated devices spanning 17 process splits, the selected HQNN, Circuit (13, 5) at L = 2, reduces overall normalized root mean square error (nRMSE) by 24.4% relative to ANN. Target-wise, the HQNN lowers $V_{th,fwd}$ RMSE from 0.297 V to 0.270 V, $V_{th,rev}$ RMSE from 0.278 V to 0.263 V, $\Delta V_{th}$ RMSE from 0.049 V to 0.045 V, SS RMSE from 22.22 mV/dec to 19.87 mV/dec, and $I_{off}$ RMSE from $5.75\times10^{-8}$ A to $4.35\times10^{-8}$ A, while $I_{on}$ RMSE remains competitive (0.053 A vs. 0.056 A). Controlled ansatz ablations further show that performance depends strongly on architecture: parameter count, depth, and two-qubit gate count correlate positively with accuracy, expressibility ($D_{KL}$) correlates negatively, and controlled-rotation entanglers outperform static controlled-NOT (CNOT)-based circuits in aggregate. A depolarizing-noise study on a representative 4-qubit circuit further suggests that comparable HQNNs may be trainable or deployable on near-term quantum hardware.

## Introduction

GaN-based high electron mobility transistors (HEMTs) are widely studied for power switching because of their wide bandgap, large critical breakdown electric field, and inherent two-dimensional electron gas (2DEG) [1]. The normally-ON behavior caused by the 2DEG is unfavorable for system integration, motivating enhancement-mode architectures. Among these, recessed-gate devices with a metal-insulator-semiconductor (MIS) stack provide direct threshold-voltage control, but they also introduce sensitivity to process variables such as recess depth, dielectric composition, and surface treatment, which no analytical model can capture without sacrificing accuracy, generality, or computational efficiency [2, 3, 4].

Machine learning has become a common approach for semiconductor device modeling, with ANNs, TCAD-augmented models, and physics-informed and generative strategies demonstrated across a wide range of device platforms and design tasks [5, 6, 7, 8, 9, 10]. Compared with simulated datasets, experimental semiconductor datasets are substantially more expensive to obtain because each labeled sample requires wafer processing, process iteration, and electrical characterization. At the same time, they provide direct information about real process-induced variability, measurement noise, and coupled fabrication effects that are difficult to capture faithfully with simulation alone. Despite these advances, classical ML approaches typically degrade on small, noisy experimental datasets where fabrication stochasticity introduces correlations that dense networks cannot efficiently capture without overfitting.

These limitations motivate complementary computational approaches. Recent advances in quantum error correction and hardware have improved the outlook for reliable quantum information processing [11, 12], making it timely to evaluate whether near-term hybrid quantum-classical algorithms can improve practical modeling tasks where classical ML is constrained by small, noisy experimental datasets.

Quantum machine learning (QML) uses superposition, entanglement, and interference to define models based on quantum states and measurements [13]. QML broadly divides into two paradigms: quantum kernel methods, which use quantum feature-map circuits to compute inner products in exponentially large Hilbert spaces [14, 15], and variational HQNNs, which train parameterized quantum circuits (PQCs) end-to-end via hybrid quantum-classical gradient descent [16, 17]. Sim et al. [18] introduced quantitative expressibility and entanglement descriptors for PQC ansatz selection, providing the framework we employ. Both branches have been demonstrated across drug discovery, finance, high-energy physics, and semiconductor defect detection [19, 20, 21, 22, 23, 24]. A longstanding barrier to practical QML has been the data-loading bottleneck, wherein encoding classical data into quantum states can negate computational advantages; recent theoretical work on quantum oracle sketching suggests one possible route to reducing this bottleneck by accessing classical data in superposition through random sampling [25], bringing provable quantum advantage on classical data closer to practical consideration.

The application of QML to semiconductor device modeling remains nascent. Recent work has established quantum kernel learning as a promising route for data-constrained semiconductor fabrication modeling. In particular, Wang et al. [26] were the first to apply QML to semiconductor device modeling, demonstrating a quantum kernel-aligned regressor (QKAR) for GaN HEMT Ohmic-contact resistance on a compact experimental dataset and showing that QML can outperform classical regressors in a small-data fabrication setting. Their study targeted a single scalar process output and relied on PCA-compressed inputs to keep kernel evaluation tractable [27]. Practical recessed-gate MIS-HEMT development requires a broader modeling capability: multiple electrical characteristics must be modeled simultaneously because threshold voltage, hysteresis, subthreshold swing, on-current, and off-state leakage are physically coupled through common process factors such as recess depth, dielectric stack, surface treatment, annealing condition, and wafer position.

These considerations motivate a variational hybrid quantum-classical neural framework that operates on the original process-defined feature vector and learns shared representations for multiple coupled device targets. Unlike quantum-kernel workflows, the proposed HQNN performs end-to-end supervised learning through a parameterized quantum circuit and a classical encoder/readout, enabling direct multi-output regression while preserving process-level feature structure. This work therefore advances QML-based semiconductor modeling from single-target quantum-kernel regression toward multi-characteristic device modeling for recessed-gate GaN MIS-HEMTs.

We develop and evaluate an HQNN on 468 experimentally fabricated devices spanning 17 process splits. The model jointly models six electrical targets from a 24-dimensional fabrication/process vector. We benchmark the selected HQNN against classical baselines and perform controlled ansatz ablations over circuit depth, parameter count, entangling gate type, and expressibility to extract design guidance for small-data semiconductor regression. This study provides evidence that variational hybrid quantum models can improve multi-characteristic modeling of experimentally fabricated MIS-HEMTs and offers a circuit-design framework for future QML-assisted semiconductor process optimization.

## Background

**Recessed-gate MIS-HEMTs and critical process effects.** Recessed-gate AlGaN/GaN MIS-HEMTs achieve enhancement-mode operation by locally thinning or removing the AlGaN barrier under the gate and inserting a gate dielectric between the metal gate and the semiconductor. The gate recess enables threshold-voltage control toward enhancement mode, and the dielectric stack suppresses gate leakage, but both together increase sensitivity to processing because the recess etch, surface preparation, and dielectric stack directly set the interface trap density and fixed charge that modulate the 2DEG and channel electrostatics [1]. As a result, modest changes in cleaning chemistry, plasma conditions, dielectric composition/thickness, and post-deposition anneal can produce correlated shifts in multiple electrical results.

**Co-optimization of key electrical characteristics.** Device performance in power switching is commonly assessed from the transfer characteristic IDS–VGS through the threshold voltage, subthreshold swing (SS), and the achievable on/off currents. In MIS-structures, gate dielectric traps that charge and discharge during a bias sweep cause the forward and reverse threshold voltages to diverge, and the hysteresis window $\Delta V_{th}$ directly quantifies that trap-induced shift. These targets are not independent: the same physical mechanisms that shift threshold can also degrade SS and alter $I_{off}$, while process-driven mobility and contact resistance changes primarily impact $I_{on}$. This coupling motivates a multi-output model rather than fitting separate single-output regressors.

**Variational quantum circuits for supervised learning.** Variational quantum algorithms combine a parameterized quantum circuit with a classical optimizer, updating circuit parameters to minimize a task loss while the quantum circuit supplies a representation that may be difficult to emulate classically at similar parameter budgets [16, 17, 13]. In supervised HQNNs, classical features are encoded into a quantum state through gates such as single-qubit rotations, the state is processed by an ansatz composed of trainable rotations and entangling operations, and measurements return expectation values that are used as features for a classical readout model. The expressive capacity and inductive bias of the ansatz depend strongly on the chosen connectivity and entangling pattern; systematic template families and quantitative descriptors such as expressibility provide a principled way to compare candidate circuits under a fixed qubit budget [18]. Expressibility is quantified by the Kullback–Leibler divergence between the circuit's output distribution and the uniform Haar-random distribution; a smaller divergence indicates a more expressible circuit.

**Hybrid training and gradient estimation.** End-to-end training of hybrid quantum-classical models requires gradients with respect to both classical parameters and quantum gate angles. While classical components can be differentiated with standard backpropagation, quantum gradients are typically obtained via the parameter-shift rule, which expresses derivatives of expectation values as differences of measurements from shifted circuits [17, 29]. In statevector simulation this can be computed exactly, whereas on hardware the same estimator must contend with finite-shot statistical noise. These considerations link circuit depth and measurement strategy directly to optimization stability and sample efficiency, and they motivate compact readouts and constrained prediction heads that place the modeling burden on the quantum representation.

## Device and Dataset

**Fabrication and process splits.** The dataset consists of 468 recessed-gate AlGaN/GaN MIS-HEMTs fabricated across 17 distinct process splits that capture diverse realistic variations encountered in enhancement-mode GaN device development. The device structure and process steps are illustrated in Fig. 1; the complete process split parameters for all 17 variants are summarized in Table 1. The 17 splits systematically vary four process axes: (i) ex-situ wet cleaning — HF-based wet clean at two concentrations (variants A–B); (ii) in-situ plasma treatment — no treatment or $N_2$, Ar, and $NH_3$ plasmas at various power levels (variants A–I); (iii) gate dielectric deposition — PEALD SiN (15 and 25 nm), RTCVD SiN (15 nm), $H_2O$-based ALD $Al_2O_3$ (15 and 25 nm), and $O_3$-based ALD $Al_2O_3$ (15 nm); and (iv) post-deposition anneal (PDA) — Forming Gas vs. $O_2$ ambient. These axes interact in coupled, non-linear ways that produce complex correlations among the electrical targets.

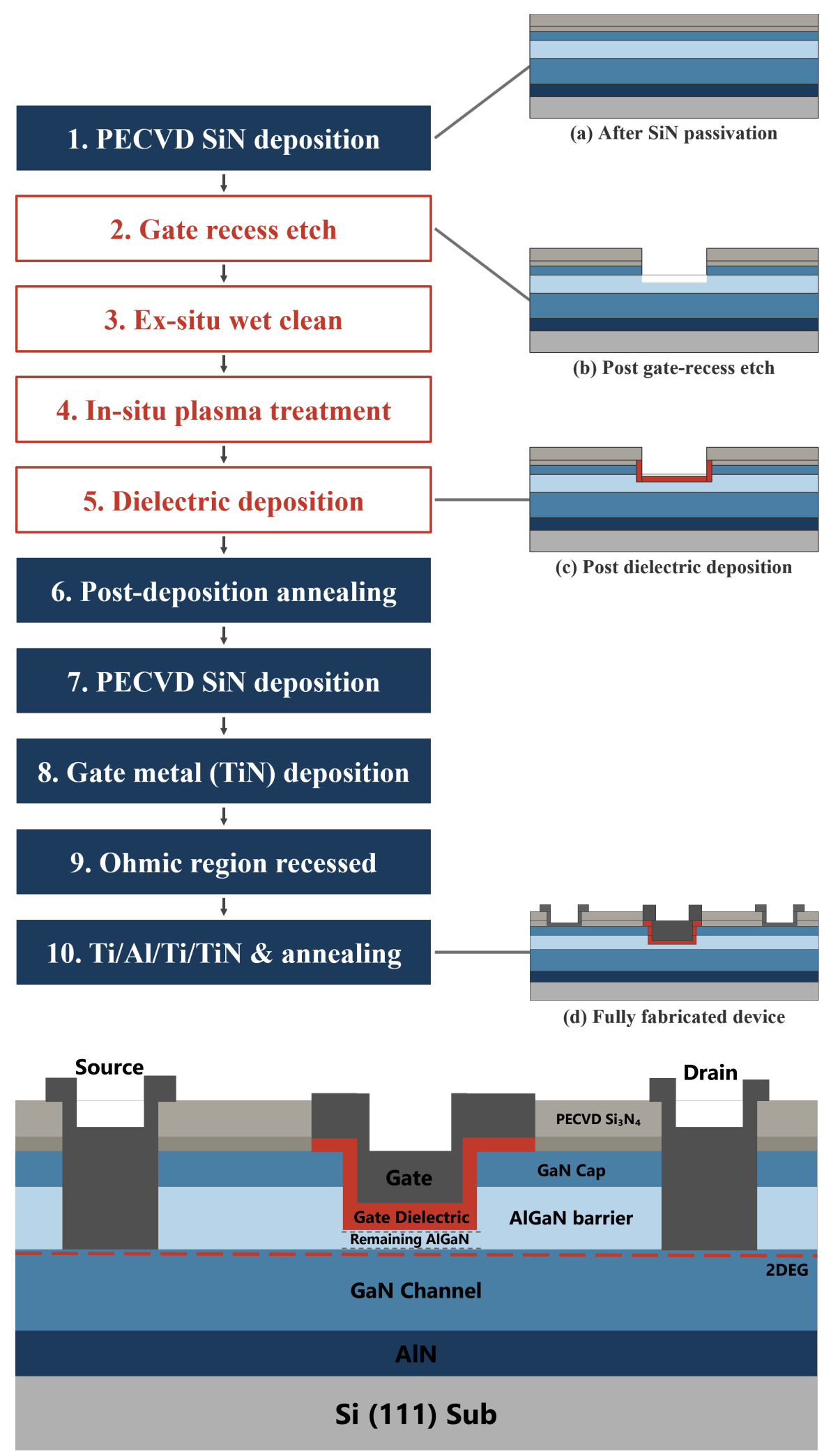

Figure 1: Recessed-gate AlGaN/GaN MIS-HEMT cross-section schematic and process flow for the fabricated devices.

Table 1: Summary of the 17 process splits (slot indices P01–P17) used in this work. All splits share a 15 nm $Al_{0.25}Ga_{0.75}N$ barrier and 3 nm GaN cap; remaining AlGaN = 18 nm - etch depth. Ex-situ wet clean variants: Clean A = HF 10%, Clean B = HF 2%. In-situ plasma variants: A = none; B = $N_2$ (20 W); C = Ar+$N_2$ (20 W); D = $N_2$ (100 W); E = Ar+$N_2$ (100 W); F = $NH_3$/Ar/$N_2$; G = $NH_3$; H = $N_2$ (50 W); I = Ar/$NH_3$. Twenty-six devices per split were characterized.

| Split | Rem. AlGaN (nm) | Ex-situ Wet | In-situ Plasma | Gate Dielectric | PDA |
|---|---|---|---|---|---|
| P01 | 13.6 | A | A | 15 nm PEALD SiN | 700°C, FG |
| P02 | 13.6 | A | B | 15 nm PEALD SiN | 700°C, FG |
| P03 | 13.6 | A | C | 15 nm PEALD SiN | 700°C, FG |
| P04 | 13.6 | A | D | 15 nm PEALD SiN | 700°C, FG |
| P05 | 13.6 | A | E | 15 nm PEALD SiN | 700°C, FG |
| P06 | 3.7 | A | A | 15 nm PEALD SiN | 700°C, FG |
| P07 | 3.7 | A | A | 25 nm PEALD SiN | 700°C, FG |
| P08 | 3.7 | A | E | 25 nm PEALD SiN | 700°C, FG |
| P09 | 5.9 | B | A | 15 nm RTCVD SiN | 700°C, FG |

| P10 | 3.7 | B | A | 15 nm RTCVD SiN | 700°C, FG |
|---|---|---|---|---|---|
| P11 | 1.5 | B | A | 15 nm RTCVD SiN | 700°C, FG |
| P12 | 3.7 | A | F | 15 nm ALD $Al_2O_3$ ($H_2O$) | 700°C, FG |
| P13 | 3.7 | A | F | 15 nm ALD $Al_2O_3$ ($H_2O$) | 500°C, $O_2$ |
| P14 | 3.7 | A | H | 15 nm ALD $Al_2O_3$ ($H_2O$) | 700°C, FG |
| P15 | 3.7 | A | G | 25 nm ALD $Al_2O_3$ ($H_2O$) | 700°C, FG |
| P16 | 3.7 | A | I | 25 nm ALD $Al_2O_3$ ($H_2O$) | 700°C, FG |
| P17 | 3.7 | A | G | 15 nm ALD $Al_2O_3$ ($O_3$) | 700°C, FG |

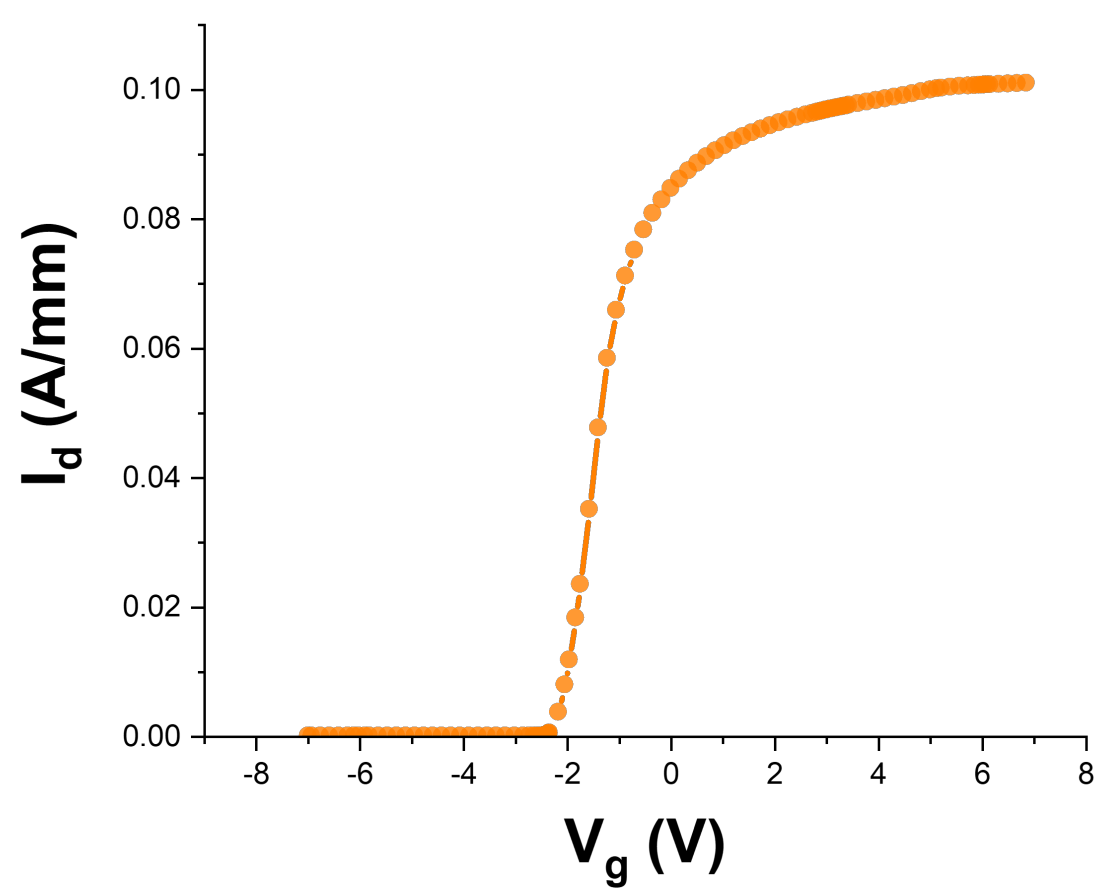

Figure 2: Typical $I_{DS}$–$V_{GS}$ transfer curve of a recessed-gate AlGaN/GaN MIS-HEMT.

**Electrical characterization.** Transfer characteristics were measured with a gate voltage (VG) sweep from -8 V to +4 V (forward) and back (reverse), yielding both the on-to-off and off-to-on branches of the IDS–VGS curve (Fig. 2). Six scalar targets are extracted per device: forward threshold voltage $V_{th,fwd}$, reverse threshold voltage $V_{th,rev}$, hysteresis $\Delta V_{th} = V_{th,rev} - V_{th,fwd}$, subthreshold swing SS, on-state drain current $I_{on}$, and off-state drain current $I_{off}$. The dataset is partitioned into training, validation, and test sets using a 60-20-20 random split. At 468 samples across a 24-dimensional input space, this dataset provides a data-constrained setting for evaluating whether structured quantum representations improve generalization over classical baselines.

## Proposed Method

### Feature Engineering

**Input vector construction.** The 24-dimensional input vector x comprises 5 continuous variables (wafer coordinates, ALE cycle count, remaining AlGaN thickness, recess depth) and 19 binary one-hot indicators encoding the categorical process choices. One-hot encoding is applied to all discrete process parameters (plasma treatment type and power, dielectric material, PDA condition) rather than integer ordinal encoding, ensuring that no artificial ordering is imposed on categorical choices that have no monotonic physical relationship.

**Spatial positional encoding.** Wafer-level spatial gradients in etch rate, temperature, and plasma exposure introduce systematic die-position dependence in device parameters. We encode die location as Cartesian coordinates (Xcoord, Ycoord) centered on the wafer. An ablation study comparing three encoding schemes (Cartesian (X, Y), polar (r, φ), and raw chip-index) showed that Cartesian coordinates produced the lowest cross-validated prediction error on threshold voltage, consistent with the dominance of linear spatial gradients (e.g., center-to-edge etch uniformity) over radial or discrete positional structure. Wafer map coordinates are shown in Fig. 3.

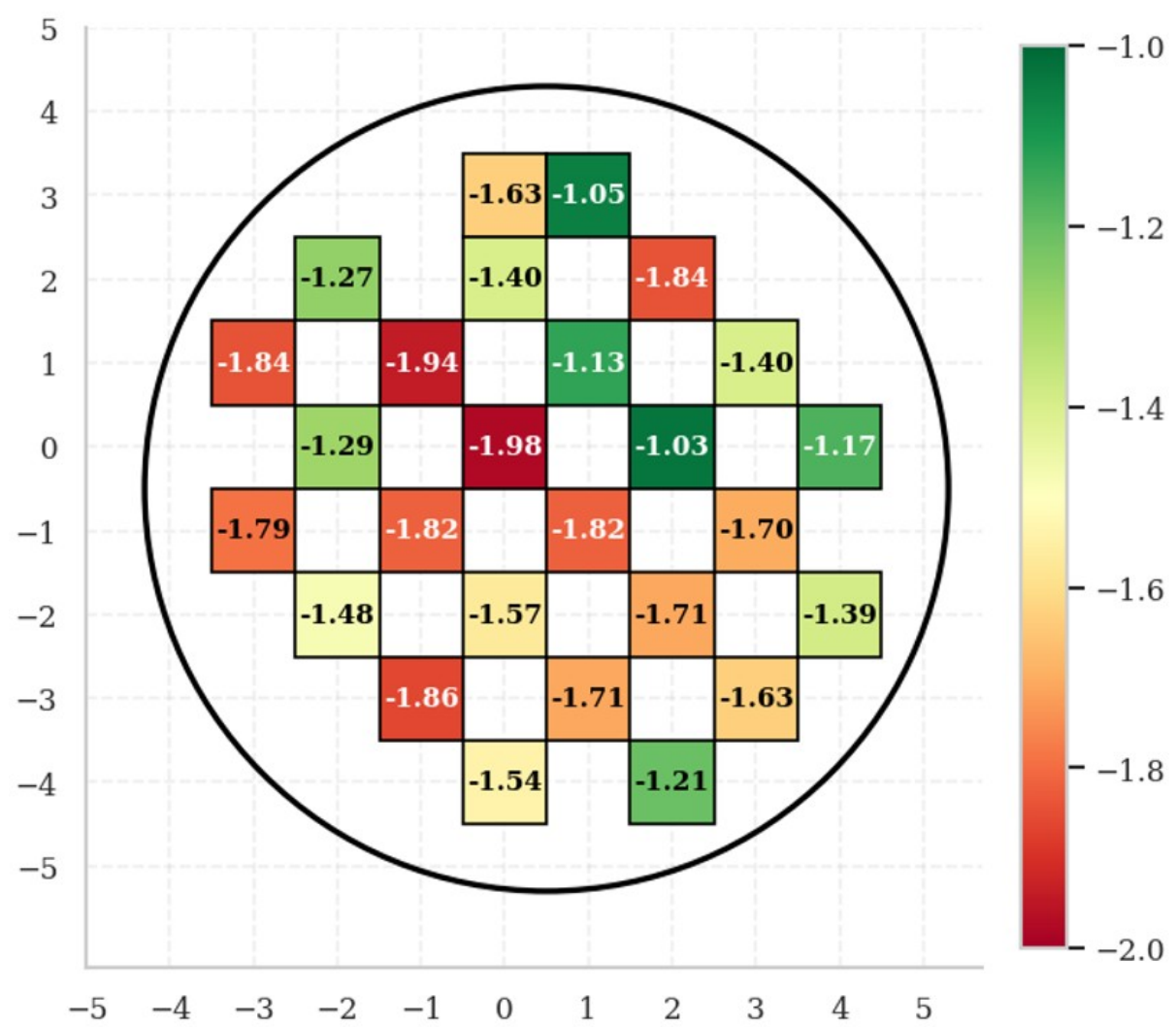

Figure 3: Wafer map showing die positions encoded as Cartesian coordinates (Xcoord, Ycoord). Wafer-level spatial gradients in etch rate, temperature, and plasma exposure introduce systematic die-position dependence in device parameters.

**No dimensionality reduction.** We deliberately omit PCA or any other linear dimensionality reduction prior to encoding. The classical encoder compresses the 24-dimensional input to a 4-dimensional latent code via a learned nonlinear MLP, preserving the full input covariance structure. This contrasts with quantum kernel methods [26], which require PCA to reduce the feature dimension and keep the kernel computation tractable [27]; our variational architecture processes the complete feature vector end-to-end, potentially avoiding the discarding of non-linear variance components that PCA may introduce in small, noisy experimental datasets [28].

## Hybrid Architecture

We study two HQNN backbones that share the same input representation, a common 19-template PQC family, and the same three-axis Pauli readout. First, a strict-bottleneck HQNN maps a 24-dimensional fabrication process vector to six device electrical targets (Fig. 4). It enforces a quantum bottleneck: a classical encoder compresses the input to one scalar per qubit, a parameterized quantum circuit (PQC) transforms that latent code, and a minimal linear readout operates on the measured quantum features with no classical bypass. This strict-bottleneck backbone is used for the single-template sweep and the descriptor-based circuit ablation analyses. Second, for sequential mixed-circuit model selection we use a dual-branch hybrid backbone in which a shared encoder produces a 16-dimensional latent code, a lightweight classical branch and a quantum branch (via a learned 16→4 projection into the PQC). The $V_{th,fwd}$, $V_{th,rev}$, $\Delta V_{th}$, and SS heads receive concatenated classical and quantum features; $I_{on}$ and $I_{off}$ are routed through quantum-only heads.

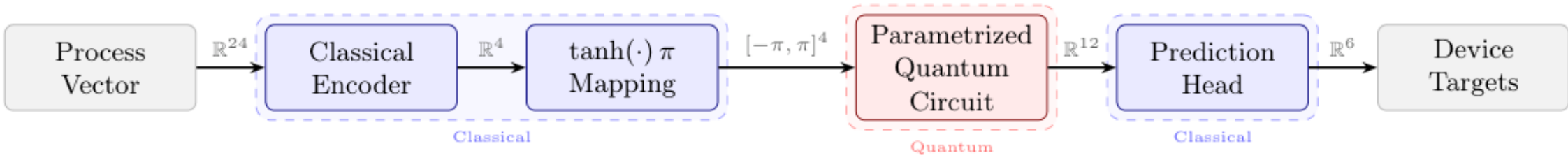

Figure 4: Strict-bottleneck HQNN backbone used for the single-template sweep and descriptor-based ablation analysis. The classical encoder compresses 24 process features to a 4-dimensional latent code, the PQC performs a structured nonlinear transformation, and a minimal linear head maps 12 Pauli expectations to 6 device targets. The dual-branch hybrid used for mixed-circuit model selection is described in the text.

**Input representation.** The input $x \in \mathbb{R}^{24}$ comprises 5 continuous process/geometry variables (wafer coordinates, ALE cycle count, remaining AlGaN thickness, recess depth) and 19 binary/one-hot process indicators (surface treatments, dielectric type/thickness, anneal condition). Continuous inputs are standardized to zero mean and unit variance; binary indicators are passed through unchanged. Six targets are predicted: $V_{th,fwd}$, $V_{th,rev}$, $\Delta V_{th}$, SS, $I_{on}$, and $I_{off}$ (log-transformed before standardization).

**Classical encoder.** A multi-layer perceptron compresses the mixed continuous-binary input to a compact latent representation. In the strict-bottleneck HQNN, the encoder maps 24 → 64 → 32 → 16 → 4 and the resulting 4-dimensional latent code is mapped to bounded rotation angles via $\theta = \pi \tanh(h)$, $\theta_i \in (-\pi, \pi)$. In the dual-branch hybrid, the shared trunk maps 24 → 64 → 32 → 16, a lightweight classical branch consumes the 16-dimensional latent code, and a learned 16→4 quantum projection produces the PQC input angles.

**Quantum feature encoding.** The circuit operates on Q = 4 qubits initialized as $|0000\rangle$. Each qubit i receives a single $RX(\theta_i)$ rotation; the ansatz, not the encoding layer, carries the main burden of nonlinear transformation.

## Variational Ansatz Design

The PQC is instantiated from a 19-template family derived from Sim et al. [18], spanning no-entanglement, nearest-neighbor chain, circuit-block, and all-to-all connectivity (Table 3). All templates share the same external interface (4 input angles, 12 output features) and differ only in internal gate structure. Every variational level carries its own independent trainable parameters. The ansatz designs are built using CUDA-Q [36], a unified programming platform that provides a hardware-agnostic, open-source programming model supporting seamless integration of quantum and classical computing resources. The two ansatz variants introduced below differ in how templates are assigned to the L levels, and they are evaluated under the strict-bottleneck (single-template) or dual-branch (mixed-circuit) backbones described in the Proposed Method section. Fig. 5 illustrates the quantum processing pipeline with a representative single-template ansatz.

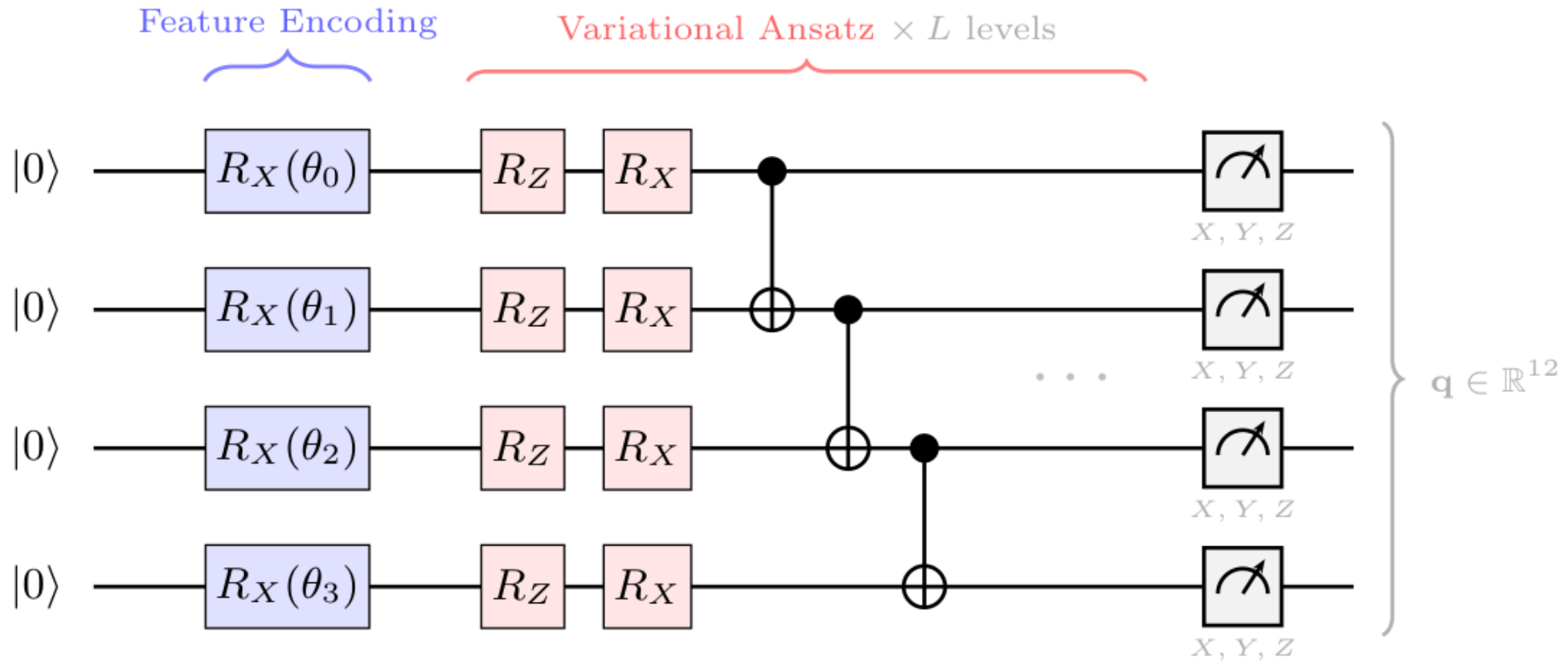

Figure 5: Quantum processing pipeline for a representative 4-qubit ansatz (Circuit 2). The encoding layer injects one learned angle per qubit; the variational ansatz is repeated for L levels; three-axis Pauli measurement yields a 12-dimensional feature vector.

The entangling gate type is a key choice: templates employ CNOT (fixed entanglement), CRZ (parameterized phase), or CRX (parameterized amplitude). Matched template pairs (e.g., 3/4, 5/6, 7/8) differ only in entangler type to enable controlled comparison. The quantum parameter count ranges from 4 (template 9, L = 1) to 140 (templates 5/6, L = 5).

The full circuit definitions for all 19 templates are listed in Table 3 (Appendix A).

### Single-Template Repeated Ansatz

In the single-template variant, a single template is drawn from the 19-template family and then repeated for L ∈ {1, ..., 5} variational levels, with independent trainable parameters at every level. End-to-end, the scaled descriptors pass through the deep encoder 24 → 64 → 32 → 16 → 4, the resulting 4-dimensional latent code is injected into a PQC whose L levels all share the same template structure, and the resulting 12-dimensional Pauli vector is mapped to the six targets by a minimal readout. Training minimizes the weighted multi-target MSE over ($V_{th,fwd}$, $V_{th,rev}$, $\Delta V_{th}$, SS, $I_{on}$, $I_{off}$). This configuration underlies the large-scale single-template sweep used for circuit ablations and design recommendations; the best single-template circuit from this sweep is shown in Fig. 8.

### Sequential Mixed-Circuit Ansatz

In the sequential mixed-circuit variant, the PQC executes an ordered sequence in which each variational level can use a different template from the same 19-template family rather than repeating a single block. We denote such a sequence by an ordered tuple (A, B), where template A fills level 1 and template B fills level 2, each carrying its own trainable parameters; the two templates need not share a gate family, entangler type, or expressibility score. Stacking heterogeneous templates this way expands the reachable subspace of the 2n-dimensional Hilbert space without increasing qubit count. In our experiments, the mixed-circuit sweeps used for model selection are evaluated with the dual-branch hybrid backbone described above (a shared encoder with a lightweight classical branch; $V_{th,fwd}$, $V_{th,rev}$, $\Delta V_{th}$, and SS heads on concatenated features; $I_{on}$ and $I_{off}$ heads on quantum features only). We exclude mixed-circuit results from the descriptor-based circuit design-insight analyses in the Circuit Ablations and Design Insights section because heterogeneous stacks do not admit a clear, physically grounded single descriptor (e.g., expressibility or entangling capability), and because the dual-branch backbone would confound circuit-only attribution.

## Readout and Training

**Three-axis measurement.** A Z-only readout from $Q = 4$ qubits provides only four expectation values, which forms a narrow readout bottleneck for predicting six coupled device targets. To expose a richer set of quantum features, we measure each qubit along the three Pauli axes:

$$q = \begin{bmatrix} \langle Z_0\rangle, \dots, \langle Z_3\rangle \\ \langle X_0\rangle, \dots, \langle X_3\rangle \\ \langle Y_0\rangle, \dots, \langle Y_3\rangle \end{bmatrix} \in \mathbb{R}^{12},$$

The resulting $3Q = 12$-dimensional feature vector captures population information through $Z$-basis measurements as well as coherence-sensitive components through the $X$ and $Y$ bases. This provides a more flexible linear readout for the six electrical targets while preserving the strict quantum bottleneck, since all target predictions are still derived only from measured PQC expectation values.

**Prediction heads.** In the strict-bottleneck HQNN, a single linear layer $\hat{y} = Wq + b$ ($W \in \mathbb{R}^6 \times 12$) maps the 12 quantum features to the 6 targets. In the dual-branch hybrid, the 16-dimensional classical latent vector and the 12-dimensional quantum feature vector are concatenated for the $V_{th,fwd}$, $V_{th,rev}$, $\Delta V_{th}$, and SS heads; the $I_{on}$ and $I_{off}$ heads operate on the quantum features alone.

**Training objective.** The model minimizes a weighted multi-target MSE:

$$\mathcal{L} = \frac{1}{T}\sum_{t=1}^{T} w_t\,(\hat{y}_t - y_t)^2,$$

**Gradient computation.** On quantum hardware, classical backpropagation cannot be applied directly through a quantum circuit, because intermediate quantum states are not accessible for gradient computation. Instead, gradients of the PQC parameters are obtained via the parameter-shift rule [17, 29]: for any trainable rotation angle $\theta_j$, the partial derivative of an expectation value $\langle O\rangle$ is computed exactly as

$$\frac{\partial\langle O\rangle}{\partial\theta_j} = \frac{1}{2}\left[\langle O\rangle|_{\theta_j+\frac{\pi}{2}} - \langle O\rangle|_{\theta_j-\frac{\pi}{2}}\right],$$

requiring two circuit evaluations per parameter per observable. Each evaluation shifts the single parameter $\theta_j$ by $\pm\pi/2$ while holding all other parameters fixed, and measures the same Pauli observable. The resulting gradient is exact, not a finite-difference approximation, for gates of the form $e^{-i\theta G/2}$ with $G^2 = I$, which includes the RX, RY, and RZ gates used in our ansatz family. In the present work, PQC gradients are computed analytically via statevector simulation, which is equivalent to the parameter-shift rule in the exact (noiseless) limit. CUDA-Q can be integrated with PyTorch-based training loops, enabling end-to-end gradients that combine classical backpropagation through the encoder and prediction head with PQC gradients via the chain rule.

## Results and Discussion

### Evaluation Setup

We report test-set root mean square error (RMSE) for each of the six targets and an overall scale-normalized score (nRMSE) aggregated across targets, defined below. We perform circuit sweeps to (i) map performance across the single-template family for ablation and design recommendations and (ii) select a mixed-circuit configuration for head-to-head comparison, with results averaged across multiple train/test splits generated from fixed seeds to assess robustness. Concretely, we sweep the 19 single-template ansatzes in Table 3 over variational depth L ∈ {1, ..., 5} and evaluate each circuit and depth setting, keeping the model initialization fixed for reproducibility. These results are later aggregated for the circuit ablation analysis in the Circuit Ablations and Design Insights section.

The aggregate nRMSE is defined as:

$$s_i = \mathrm{IQR}(y_{i,\mathrm{train}}) = Q_{0.75}(y_{i,\mathrm{train}}) - Q_{0.25}(y_{i,\mathrm{train}})$$

$$\mathrm{nRMSE}_i = \frac{\mathrm{RMSE}_i}{s_i}$$

$$\text{Overall nRMSE} = \frac{1}{T}\sum_{i=1}^{T}\mathrm{nRMSE}_i,$$

nRMSE is used in place of averaged raw RMSE, which mixes incomparable units and scales across targets, and in place of MAPE, which is unstable near zero (notably for $I_{off}$).

The single-template sweep is used primarily for circuit ablations and design recommendations; it quantifies how performance varies with template choice and depth under a controlled strict-bottleneck backbone. For model selection and head-to-head comparisons, we run an analogous sweep for the sequential mixed-circuit variant by evaluating all ordered template pairs at fixed depth L = 2 under the dual-branch hybrid backbone described in the Proposed Method section. As the classical reference point, we use an ANN baseline trained and evaluated identically.

### Performance Against Classical Baselines

The best single-template circuit (template 13, L = 4, circuit-block CRZ connectivity; Fig. 8(a)) repeats the following unit four times with independent parameters:

$$\begin{aligned}U_k(\theta_k) = {} & [CR_Z(2,1)\cdot CR_Z(3,2)\cdot CR_Z(0,3)\cdot CR_Z(1,0)] \\ & \cdot \otimes_{i=0}^{3} R_Y^{(i)}(\theta)\cdot[CR_Z(0,1)\cdot CR_Z(1,2) \\ & \cdot CR_Z(2,3)\cdot CR_Z(3,0)]\cdot\otimes_{i=0}^{3} R_Y^{(i)}(\theta),\end{aligned}$$

giving $U = U_4U_3U_2U_1$. The best mixed-circuit (Circuit (13, 5), L = 2; Fig. 8(b)) reuses template 13 (circuit-block CRZ) at level 1, so $U_1(\theta_1)$ has the structure of the unit shown above, and appends template 5 (all-to-all CRZ) at level 2:

$$\begin{aligned}U_2(\theta_2) = {} & \otimes_{i=0}^{3}(R_Z^{(i)}R_X^{(i)})\cdot[CR_Z(2,3)\cdot CR_Z(1,3) \\ & \cdot CR_Z(1,2)\cdot CR_Z(0,3)\cdot CR_Z(0,2) \\ & \cdot CR_Z(0,1)]\cdot\otimes_{i=0}^{3}(R_Z^{(i)}R_X^{(i)}),\end{aligned}$$

giving $U = U_2(\theta_2)\,U_1(\theta_1)$. Gate order is right-to-left; each $U_k(\theta_k)$ denotes one independent variational level.

From the mixed-circuit sweep we select Circuit (13, 5) (L = 2) as the HQNN used for head-to-head comparison. Across the validation splits, this circuit ranks first in overall nRMSE. The strict-bottleneck single-template sweep is retained for circuit ablations and design recommendations; its best single-template circuit (template 13 at L = 4) is shown in Fig. 8(a), and the best mixed-circuit (Circuit (13, 5) at L = 2) is shown in Fig. 8(b).

For the classical comparison, we benchmark the selected HQNN against ANN, PLS, DecisionTree, MTL-ElasticNet, KernelRidge, and XGBoost. ANN is the primary reference because neural-network models are common baselines in prior semiconductor-device studies and, within this set, provide the strongest classical performance. The additional baselines provide coverage of linear, tree-based, and kernel-based model families. Because several of these baselines treat each target independently or only weakly couple the outputs, the HQNN-versus-ANN comparison is the most relevant practical reference for evaluating joint multi-target modeling.

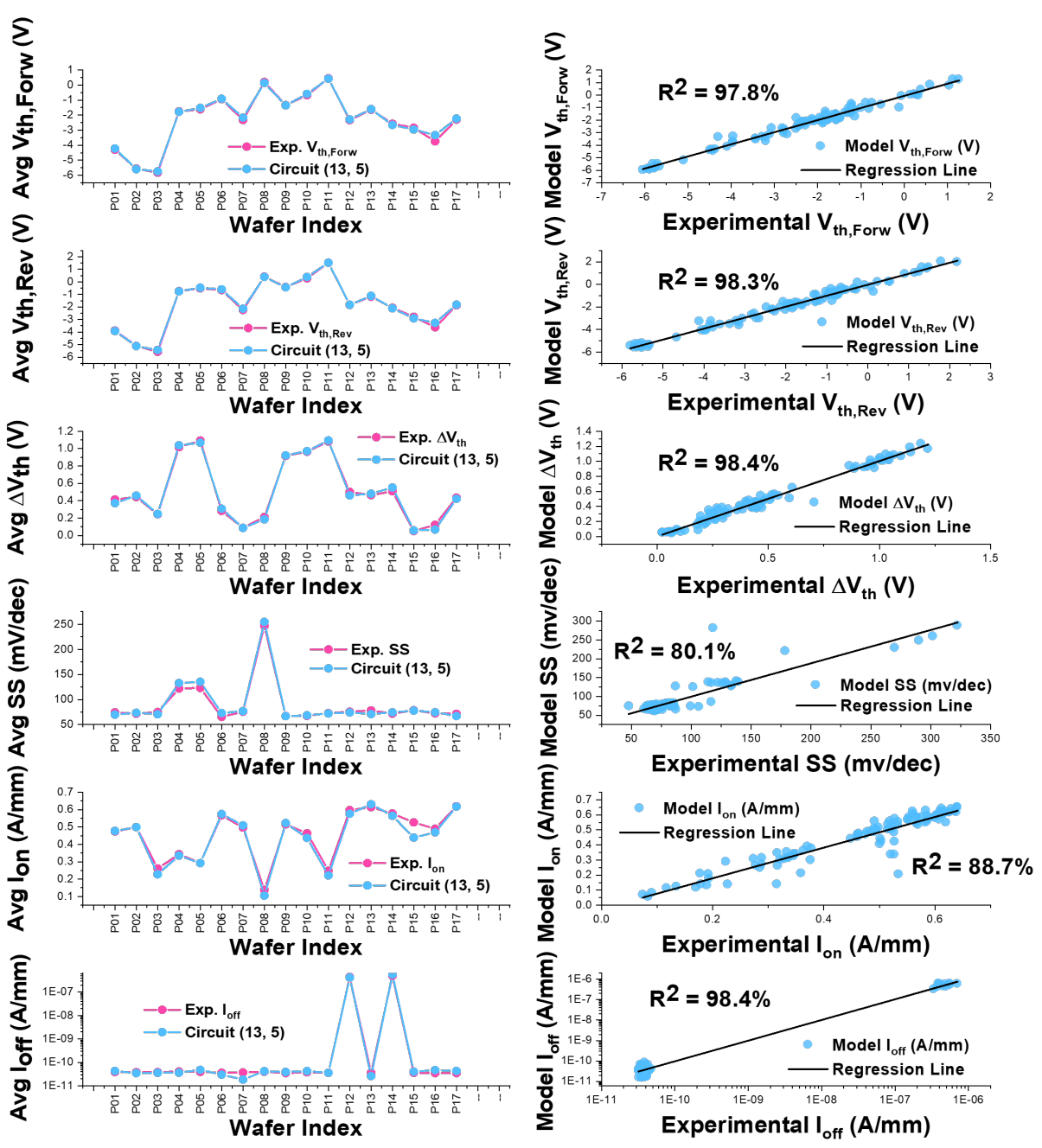

Figure 6: Wafer-level means (left) and global parity plots (right) for the selected HQNN configuration: Circuit (13, 5) (L = 2); overall $R^2$ = 0.9229.

Among the tested baselines, Circuit (13, 5) yields the lowest aggregate error, reducing overall nRMSE from 2395.6 for the ANN to 1811.4, a ~24% reduction. As summarized visually in Fig. 7, the HQNN yields the lowest error on the overall metric and on five of the six individual targets. The voltage-like targets and SS also improve consistently, with RMSE reductions of 0.026 V for $V_{th,fwd}$ (0.297 V→0.270 V), 0.015 V for $V_{th,rev}$ (0.278 V→0.263 V), 0.004 V for $\Delta V_{th}$ (0.049 V→0.045 V), and 2.35 mV/dec for SS (22.22 mV/dec→19.87 mV/dec), while $I_{off}$ RMSE decreases by approximately 24% ($5.75\times10^{-8}$ A→$4.35\times10^{-8}$ A). The only target on which the ANN remains slightly better is $I_{on}$ (0.053 A vs. 0.056 A), indicating a localized trade-off rather than a reversal of the overall ranking.

The improvement in $I_{off}$ is practically relevant because off-state leakage is among the noisiest and most process-sensitive targets in the dataset. It is influenced by coupled stochastic mechanisms, including interface-trap dynamics, plasma-induced surface damage, dielectric fixed-charge fluctuations, and leakage-path variability, which introduce nonlinear variation that is difficult to learn from limited experimental data. The concurrent gains in SS and the voltage-related outputs indicate that the improvement is not confined to a single leakage metric. In device optimization, inaccurate leakage prediction can favor designs that appear attractive by voltage or current criteria but fail off-state requirements.

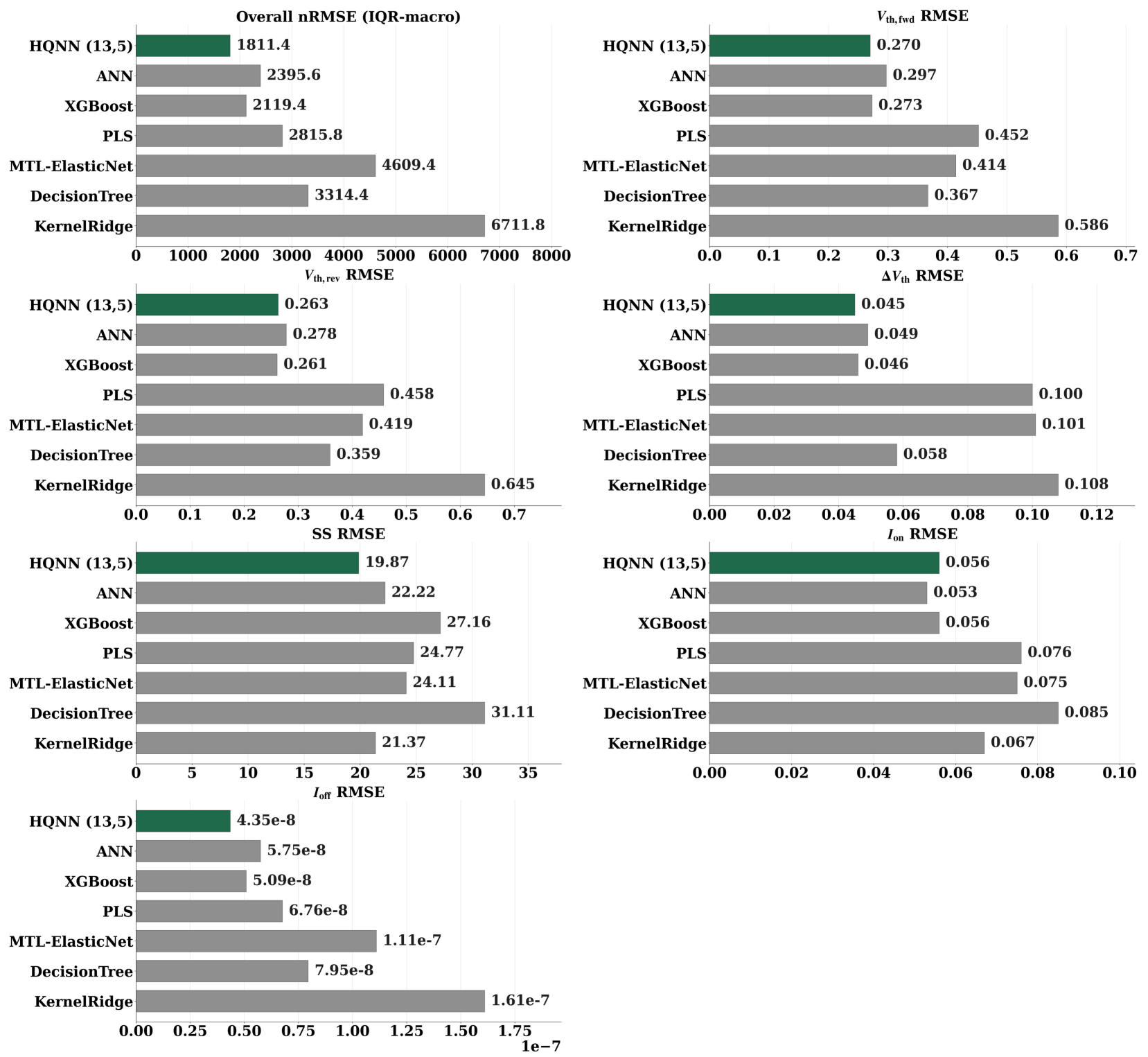

Figure 7: Comparison of overall nRMSE (IQR-macro) and per-target raw RMSE for the selected HQNN and the classical baselines. Lower values indicate better predictive accuracy.

## Circuit Ablations and Design Insights

To interpret the performance differences across ansatz choices, we analyze which circuit properties are most predictive of accuracy at the level of split-averaged single-template circuit and level configurations. The sequential mixed-circuit results are excluded, because for heterogeneous template stacks there is no clear, physically grounded way to assign a single expressibility score, entangling-capability score, or related higher-order descriptor to the composite circuit.

Across these configurations, overall accuracy correlates negatively with expressibility D_KL ($\rho = -0.512$, $p = 1.1 \times 10^{-7}$) and positively with parameter count ($\rho = +0.426$, $p = 1.7 \times 10^{-5}$), circuit depth ($\rho = +0.335$, $p = 9.2 \times 10^{-4}$), and two-qubit gate count ($\rho = +0.317$, $p = 1.7 \times 10^{-3}$). Because lower D_KL denotes a more expressive circuit, the negative sign indicates that more expressible circuits achieve higher accuracy on average. At the extreme, however, the very most expressible circuits are not the top performers. This is consistent with barren plateaus [30]: circuits too close to Haar-random have exponentially flat gradient landscapes, making reliable optimization progressively harder in a small-data regime.

Gate family also affects performance, in line with the analysis of Sim et al. [18]: CNOT-based circuits underperform their CRZ and CRX counterparts. This is expected: relative to static CNOT entanglers, parameterized controlled rotations provide additional tunable degrees of freedom, giving the optimizer more expressive capacity without increasing circuit depth. Both the best single-template circuit and the best mixed-circuit use exclusively CRZ or CRX entanglers.

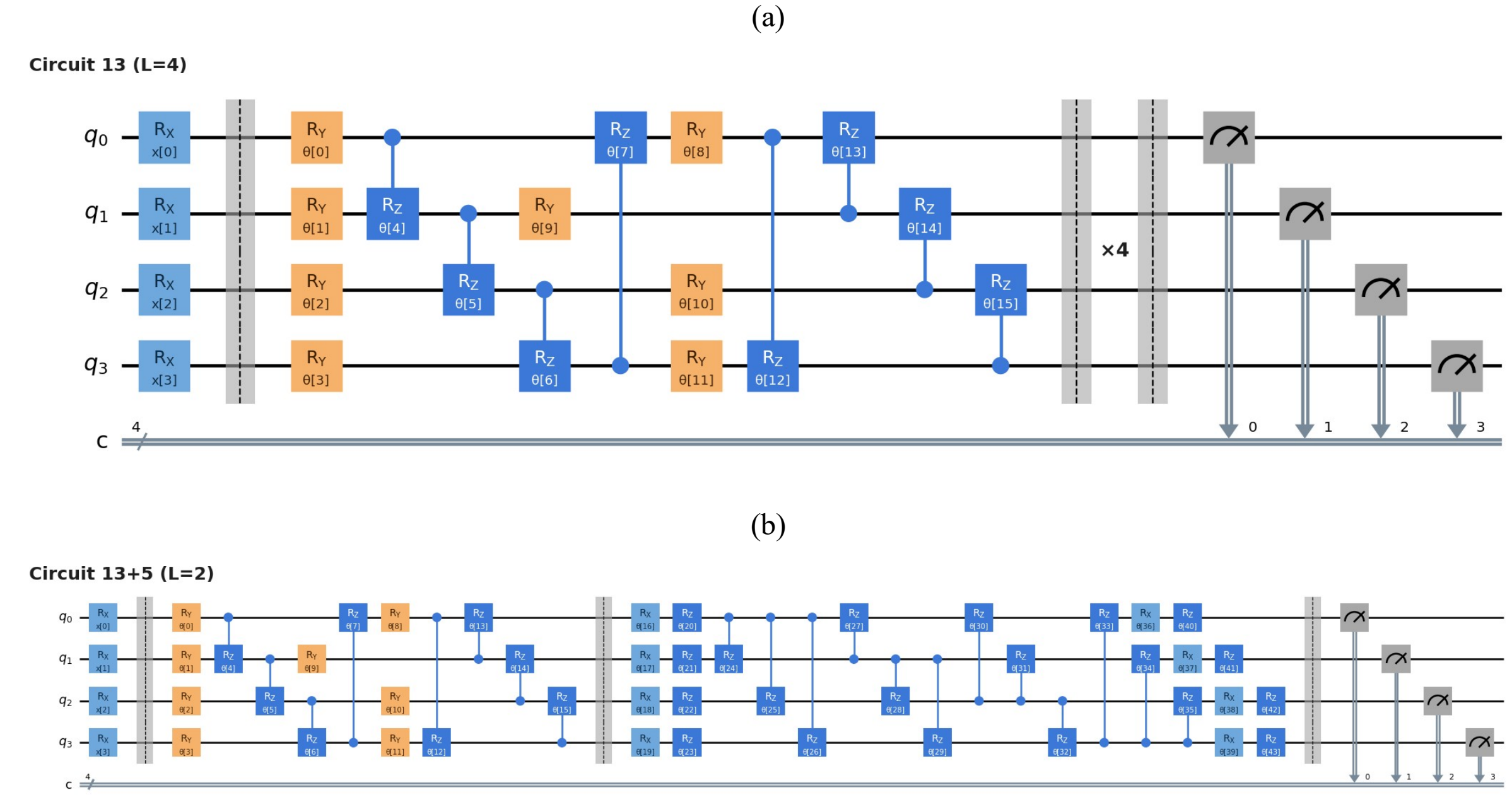

Figure 8: Circuit diagrams of (a) the best single-template circuit from the single-template sweep (template 13, L = 4) and (b) the best mixed-circuit configuration selected for head-to-head comparison (Circuit (13, 5), L = 2).

## Noise Robustness Under Depolarizing Errors

To probe sensitivity to gate-level noise, we retrained Circuit 13 (L = 4) under a depolarizing noise model applied after each single- and two-qubit gate. Table 2 reports overall and per-target $R^2$ at the tested noise levels, and Fig. 9 highlights the $I_{off}$ sensitivity via $\Delta R^2$ relative to the clean baseline.

While depolarizing noise is an idealized stochastic Pauli channel, it provides a compact effective-noise model for studying how gate-level errors may affect circuit performance after complex hardware errors are summarized through twirling- and benchmarking-based characterizations such as randomized benchmarking [31, 32]. The tested probabilities were chosen to span plausible effective error regimes rather than to model a specific device: $p = 0.005$ (0.5%) represents a sub-percent per-gate depolarization setting comparable to selected high-fidelity superconducting and fluxonium gate reports [33, 34, 35]. Operationally, $p = 0.005$ corresponds to roughly one stochastic error opportunity per ~200 gates, whereas $p = 0.05$ corresponds to ~1 per 20 gates. $p = 0.05$ (5%) is intentionally pessimistic and serves as an upper stress-test beyond typical high-fidelity device-operation regimes.

At $p_1 = p_2 = 0$, the model achieves overall $R^2 = 0.903$. Mild depolarizing noise ($p_1 = p_2 = 0.005$) yields a comparable overall score ($R^2 = 0.905$). As $p_1$ increases to 0.05 (with $p_2 = 0.005$ fixed), overall performance declines to 0.853, driven primarily by $I_{off}$, which drops from 0.901 to 0.601 ($\Delta R^2$ = - 0.300). Voltage-related targets and SS remain comparatively stable across the tested range, consistent with those targets being well-approximated by the classical encoder alone, so that degradation of the quantum representation has limited impact on their predictions. At higher noise levels, strongly depolarized qubit states approach a maximally mixed distribution, so the Pauli expectation values measured by the readout layer tend toward zero and carry diminishing task-relevant information; the prediction head must therefore rely increasingly on the residual structure encoded by the classical encoder rather than on the quantum representation, effectively reducing the HQNN to a degraded classical model at the $p_1 = 0.05$ extreme.

Table 2: Noise robustness of Circuit 13 (L = 4) under depolarizing errors. $p_1$: single-qubit depolarizing probability; $p_2$: two-qubit depolarizing probability. Overall $R^2$ is the arithmetic mean of the six target scores.

| $p_1$ | $p_2$ | Overall | $V_{th,fwd}$ | $V_{th,rev}$ | $\Delta V_{th}$ | SS | $I_{on}$ | $I_{off}$ |
|---|---|---|---|---|---|---|---|---|
| 0.000 | 0.000 | 0.903 | 0.964 | 0.971 | 0.966 | 0.751 | 0.867 | 0.901 |
| 0.005 | 0.005 | 0.905 | 0.967 | 0.973 | 0.963 | 0.780 | 0.896 | 0.849 |
| 0.010 | 0.005 | 0.883 | 0.971 | 0.973 | 0.962 | 0.767 | 0.878 | 0.748 |
| 0.050 | 0.005 | 0.853 | 0.953 | 0.960 | 0.953 | 0.780 | 0.871 | 0.601 |

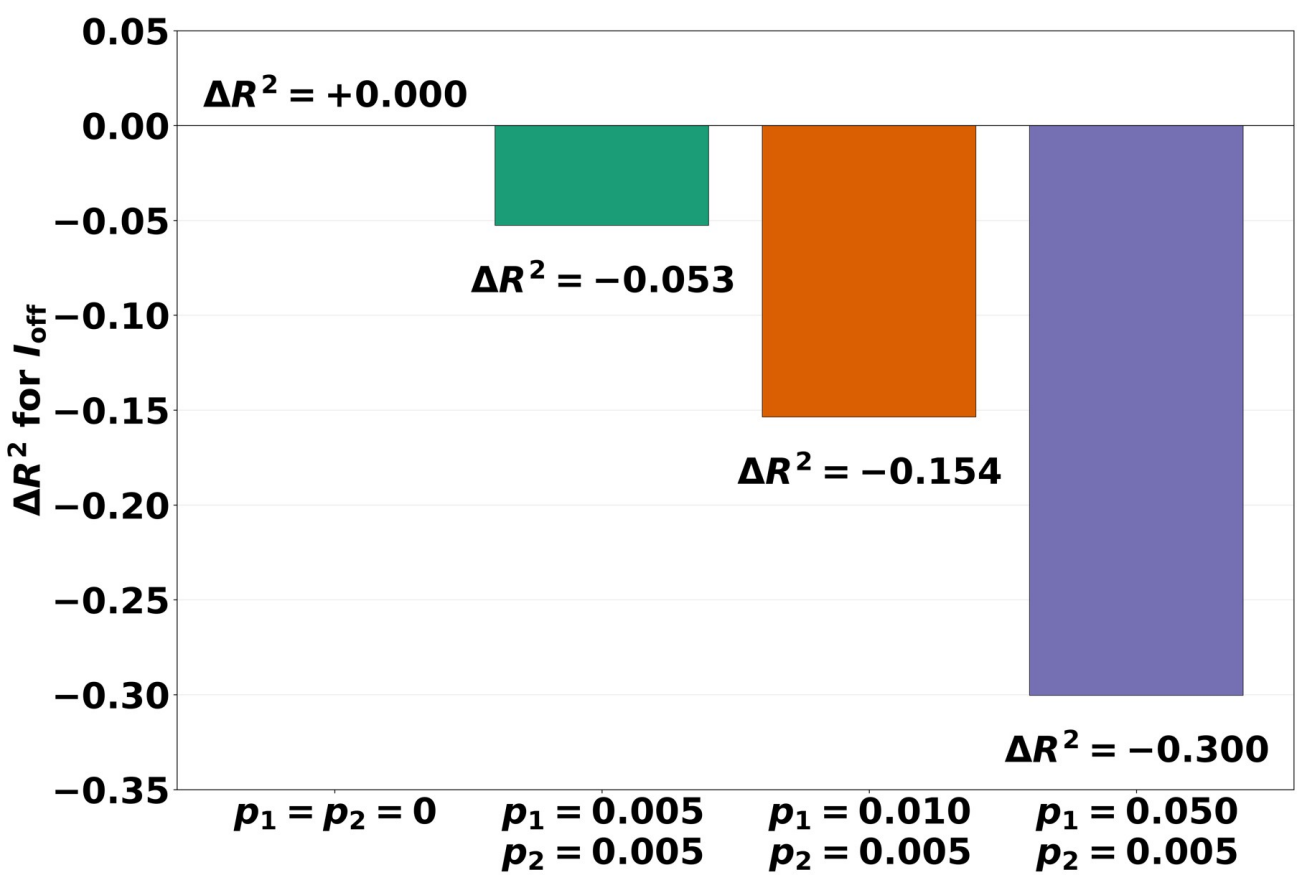

Figure 9: $I_{off}$ $\Delta R^2$ under depolarizing noise relative to the clean baseline ($p_1 = p_2 = 0$). Each bar is labeled with its $\Delta R^2$ value.

## Conclusion

We presented a multi-target HQNN for multi-characteristic modeling of recessed-gate AlGaN/GaN MIS-HEMTs and performed large-scale circuit sweeps to extract circuit-design guidance and select a mixed-circuit HQNN for head-to-head comparison. On the 468-device experimental dataset, the selected HQNN, Circuit (13, 5) at L = 2, reduces overall nRMSE by 24.4% relative to ANN. The most pronounced gains appear on the practically difficult targets, with SS RMSE falling from 22.22 mV/dec to 19.87 mV/dec and Ioff RMSE falling from $5.75\times10^{-8}$ A to $4.35\times10^{-8}$ A (a ~24% reduction). The HQNN also lowers $V_{th,fwd}$ RMSE from 0.297 V to 0.270 V, $V_{th,rev}$ RMSE from 0.278 V to 0.263 V, and $\Delta V_{th}$ RMSE from 0.049 V to 0.045 V, while $I_{on}$ RMSE remains competitive (0.053 A vs. 0.056 A). Beyond the aggregate accuracy gain, the single-template circuit ablations provide practical design guidance for similar small-data, noisy regression tasks. Across the ansatz sweep, performance improves with parameter count, circuit depth, and two-qubit gate count, but declines as circuits approach overly expressive Haar-like behavior, consistent with the negative correlation between accuracy and expressibility. Parameterized controlled-rotation entanglers also outperform static CNOT-based circuits in aggregate, indicating that trainable entangling structure is more effective for this experimental semiconductor regression setting. A depolarizing-noise study on a representative strict-bottleneck 4-qubit circuit (template 13, L = 4) shows that the model retains overall $R^2 = 0.905$ at mild noise ($p_1 = p_2 = 0.005$), with $I_{off}$ remaining the most noise-sensitive target and the voltage-like outputs staying largely stable. Taken together, these results support the feasibility of training or deploying comparable HQNNs on near-term real 4-qubit quantum registers and highlight HQNNs as effective models for MIS-HEMT co-optimization when experimental data are expensive yet uniquely informative about real fabrication variability.

## Acknowledgements

This work was financially supported by the “Advanced Semiconductor Technology Research Center” from The Featured Areas Research Center Program within the framework of the Higher Education Sprout Project by the Ministry of Education (MOE) in Taiwan, and is also supported by the National Science and Technology Council (NSTC), Taiwan, under Grant 112-2628-E-A49-020-MY3 and 114-2223-E-A49-003-MY4. The authors thank NVIDIA for support through the NVIDIA Academic Grant Program, technical assistance and GPU computing resources from NVIDIA AI Technology Center (NVAITC), CUDA-Q support, and valuable scientific discussions with Chi-Cheng Fu. We also thank the National Center for High-Performance Computing (NCHC) for helpful discussions and for providing computing resources.

## Appendix A

Table 3: Variational ansatz template inventory for Q = 4 qubits [18]. $R_X^{(i)}$, $R_Y^{(i)}$, $R_Z^{(i)}$, $H^{(i)}$: single-qubit gates on qubit i. CX(c,t), CZ(c,t): fixed entanglers. $CR_X$(c,t), $CR_Z$(c,t): parameterized entanglers (c = control, t = target). $\bigotimes_i$ ranges over all four qubits unless bounded explicitly. Gate order is right-to-left.

| ID | Connectivity | Circuit unit layer $U_k(\theta)$ |
|---|---|---|
| 1 | Nearest-neighbor (local) | $\bigotimes_{i=0}^{3}(R_Z^{(i)} R_X^{(i)})$ |
| 2 | Nearest-neighbor | $[CX(2,3) \cdot CX(1,2) \cdot CX(0,1)] \cdot \bigotimes_i (R_Z^{(i)} R_X^{(i)})$ |
| 3 | Nearest-neighbor | $[CR_Z(2,3) \cdot CR_Z(1,2) \cdot CR_Z(0,1)] \cdot \bigotimes_i (R_Z^{(i)} R_X^{(i)})$ |
| 4 | Nearest-neighbor | $[CR_X(2,3) \cdot CR_X(1,2) \cdot CR_X(0,1)] \cdot \bigotimes_i (R_Z^{(i)} R_X^{(i)})$ |
| 5 | All-to-all | $\bigotimes_i (R_Z^{(i)} R_X^{(i)}) \cdot [CR_Z(2,3) \cdot CR_Z(1,3) \cdot CR_Z(1,2) \cdot CR_Z(0,3) \cdot CR_Z(0,2) \cdot CR_Z(0,1)] \cdot \bigotimes_i (R_Z^{(i)} R_X^{(i)})$ |
| 6 | All-to-all | $\bigotimes_i (R_Z^{(i)} R_X^{(i)}) \cdot [CR_X(2,3) \cdot CR_X(1,3) \cdot CR_X(1,2) \cdot CR_X(0,3) \cdot CR_X(0,2) \cdot CR_X(0,1)] \cdot \bigotimes_i (R_Z^{(i)} R_X^{(i)})$ |
| 7 | Nearest-neighbor | $\bigotimes_{i=1}^{2}(R_Z^{(i)} R_X^{(i)}) \cdot CR_Z(1,2) \cdot \bigotimes_{i=1}^{2}(R_Z^{(i)} R_X^{(i)}) \cdot [CR_Z(2,3) \cdot CR_Z(0,1)] \cdot \bigotimes_{i=0}^{3}(R_Z^{(i)} R_X^{(i)})$ |
| 8 | Nearest-neighbor | $\bigotimes_{i=1}^{2}(R_Z^{(i)} R_X^{(i)}) \cdot CR_X(1,2) \cdot \bigotimes_{i=1}^{2}(R_Z^{(i)} R_X^{(i)}) \cdot [CR_X(2,3) \cdot CR_X(0,1)] \cdot \bigotimes_{i=0}^{3}(R_Z^{(i)} R_X^{(i)})$ |
| 9 | Nearest-neighbor | $\bigotimes_i R_X^{(i)} \cdot CX(1,2) \cdot [CX(2,3) \cdot CX(0,1)] \cdot \bigotimes_i H^{(i)}$ |
| 10 | Ring topology | $[R_Y^{(3)} \otimes R_Y^{(0)}] \cdot [CZ(0,3) \cdot CZ(1,2)] \cdot [R_Y^{(2)} \otimes R_Y^{(1)}] \cdot [CZ(2,3) \cdot CZ(0,1)] \cdot \bigotimes_i R_Y^{(i)}$ |
| 11 | Nearest-neighbor | $CX(2,1) \cdot \bigotimes_{i=1}^{2}(R_Z^{(i)} R_Y^{(i)}) \cdot [CX(3,2) \cdot CX(1,0)] \cdot \bigotimes_{i=0}^{3}(R_Z^{(i)} R_Y^{(i)})$ |
| 12 | Nearest-neighbor | $CZ(1,2) \cdot \bigotimes_{i=1}^{2}(R_Z^{(i)} R_Y^{(i)}) \cdot [CZ(2,3) \cdot CZ(0,1)] \cdot \bigotimes_{i=0}^{3}(R_Z^{(i)} R_Y^{(i)})$ |
| 13 | Circuit-block | $[CR_Z(2,1) \cdot CR_Z(3,2) \cdot CR_Z(0,3) \cdot CR_Z(1,0)] \cdot \bigotimes_i R_Y^{(i)} \cdot [CR_Z(0,1) \cdot CR_Z(1,2) \cdot CR_Z(2,3) \cdot CR_Z(3,0)] \cdot \bigotimes_i R_Y^{(i)}$ |
| 14 | Circuit-block | $[CR_X(2,1) \cdot CR_X(3,2) \cdot CR_X(0,3) \cdot CR_X(1,0)] \cdot \bigotimes_i R_Y^{(i)} \cdot [CR_X(0,1) \cdot CR_X(1,2) \cdot CR_X(2,3) \cdot CR_X(3,0)] \cdot \bigotimes_i R_Y^{(i)}$ |
| 15 | Circuit-block | $[CX(2,1) \cdot CX(3,2) \cdot CX(0,3) \cdot CX(1,0)] \cdot \bigotimes_i R_Y^{(i)} \cdot [CX(0,1) \cdot CX(1,2) \cdot CX(2,3) \cdot CX(3,0)] \cdot \bigotimes_i R_Y^{(i)}$ |
| 16 | Nearest-neighbor | $CR_Z(1,2) \cdot [CR_Z(2,3) \cdot CR_Z(0,1)] \cdot \bigotimes_i (R_Z^{(i)} R_X^{(i)})$ |
| 17 | Nearest-neighbor | $CR_X(1,2) \cdot [CR_X(2,3) \cdot CR_X(0,1)] \cdot \bigotimes_i (R_Z^{(i)} R_X^{(i)})$ |
| 18 | Ring topology | $[CR_Z(0,1) \cdot CR_Z(1,2) \cdot CR_Z(2,3) \cdot CR_Z(3,0)] \cdot \bigotimes_i (R_Z^{(i)} R_X^{(i)})$ |
| 19 | Ring topology | $[CR_X(0,1) \cdot CR_X(1,2) \cdot CR_X(2,3) \cdot CR_X(3,0)] \cdot \bigotimes_i (R_Z^{(i)} R_X^{(i)})$ |